\documentclass[11pt,onecolumn]{article}
\usepackage{amsfonts}
\usepackage{pifont}
\usepackage{algpseudocode}
\usepackage{algorithm} 
\usepackage{caption}
\usepackage{multirow}
\usepackage{amssymb}
\usepackage{amsmath}
\usepackage{multirow}
\usepackage{makecell}
\usepackage{pdflscape}
\usepackage{geometry}
\usepackage{cases}
\usepackage{CJK}

\usepackage{caption}
\usepackage{latexsym}
\usepackage{graphicx}
\usepackage{epsfig}

\normalsize \makeatletter
\newenvironment{proof}{{\em Proof.} }{\hfill$\Box$\vspace{0.1in}}

\newtheorem{lemma}{Lemma}[section]

\newtheorem{theorem}{Theorem}[section]

\newcommand{\head}[1]
{\markright{\hbox to 0pt{\vtop to 0pt{\hbox{}\vskip 3mm \hrule
width \textwidth \vss} \hss}{\sc #1}}}

\begin{document}
\title{Improved Lower Bounds for Online Parallel Machine Scheduling}
\author{Junwen XIAO\thanks{School of Mathematical Science, Zhejiang University, Hangzhou 310027, P. R. China.} \and Zhiyi TAN\thanks{Corresponding author. School of Mathematical Science, Zhejiang University,
Hangzhou 310027, P. R. China. Supported by the National Natural
Science Foundation of China (12571340). {\tt
tanzy@zju.edu.cn}} }
\date{}
\maketitle

\begin{abstract}
This paper establishes new lower bounds for online parallel machine scheduling aimed at minimizing the makespan. By leveraging bin packing algorithms to estimate the optimal makespan, we prove that the competitive ratio of any online algorithm is at least 
$1.887111$. Additionally, via a direct construction, we demonstrate a lower bound of 
$1.74916$ for the case of five machines. Our results improve upon the best known lower bounds established two decades ago.
\vskip 2mm\noindent\textbf{Keywords:} Scheduling, Online algorithm, Competitive ratio, Parallel machines.
\end{abstract}

\section{Introduction}
Parallel machine scheduling serves as a benchmark problem in online optimization. Consider a set of $n$ independent jobs $\mathcal{J} =\{J_1, J_2, \dots, J_n\}$ to be processed on $m\geq 2$ parallel identical machines, denoted by $M_1, \ldots, M_m$. All jobs are available at time $0$ and arrive in a
given order. Upon the arrival of a job $J_j$, only the processing times of $J_j$ and its predecessors $J_1, \dots, J_{j-1}$ are revealed. Information regarding subsequent jobs, including their processing times and even their existence, remains unknown. Once a job is assigned to a machine, the decision is irrevocable. 

A schedule is formed by assigning each job to exactly one machine. The load of a machine is defined as the sum of the processing times of the jobs assigned to it, and the makespan is the maximum load over all machines. The fundamental objective in parallel machine scheduling is to minimize the makespan.

An algorithm designed for online problems is called an \emph{online algorithm}. The performance of an online algorithm $A$ is typically evaluated by its \emph{competitive ratio}, defined as the smallest number $\theta$ such that $C^A(\mathcal{J}) \leq \theta C^*(\mathcal{J})$ holds for every instance $\mathcal{J}$ \cite{ST85}. Here, $C^A(\mathcal{J})$ denotes the objective value of the solution produced by $A$, and $C^*(\mathcal{J})$ denotes the optimal objective value of the corresponding offline problem. An online problem is said to have a \emph{lower bound} $\theta$ if no online algorithm can achieve a competitive ratio strictly smaller than $\theta$. An online algorithm is called \emph{optimal} if its competitive ratio matches the lower bound. The primary goal in the study of online problems is to design algorithms with the smallest possible competitive ratios and to establish lower bounds as large as possible, until the optimal competitive ratio is determined.

Establishing lower bounds for an online problem typically involves constructing a family of adversarial instances such that no single online algorithm can perform well on all of them. This requires comparing the objective value of the solution produced by the algorithm against the offline optimum, a crucial step that, for our problem, reduces to computing the optimal makespan.
Since computing the exact optimal makespan is computationally intractable, it suffices to derive a tight upper bound on it. Previous approaches have generally derived these bounds by systematically constructing feasible schedules. To facilitate this analysis, the instance construction is kept relatively simple. A common strategy involves partitioning jobs into several phases, each containing a restricted set of job types. The processing times are carefully chosen to constrain the online algorithm, forcing it to process jobs according to specific rules in each phase.

In this paper, we establish new lower bounds for online parallel machine scheduling with a fixed number of machines, leveraging a novel approach combined with computational search. While maintaining the traditional phase-based structure, we incorporate a bin-packing strategy to derive tighter estimates of the optimal makespan. This enhancement accommodates a more diverse set of job types, yielding bounds closer to the exact optimum. We explicitly construct instances to prove lower bounds for $m \geq 6$,
with the largest bound of $1.887111$ attained at $m = 8192$. Furthermore,
for $m = 5$, we present a new lower bound of $1.74916$ via a direct construction. The resulting instance is simple enough to be verified by hand. 

Table~\ref{tab:comparison} summarizes our new lower bounds in comparison with the existing results in \cite{CVW94, Rudin01}. 
The improvements yielded by our approach are also explicitly shown.

\begin{table}[ht]
	\centering
	\small
	\setlength{\tabcolsep}{8pt}
	\renewcommand{\arraystretch}{1.18}
	\begin{tabular}{|c|c|c|c|c|c|c|c|}
		\hline
		$m$ & 5 & 6 & 7 & 8 & 9 & 10 & 11 \\
		\hline
		Lower bound in \cite{CVW94}  & 1.746250 & 1.773010 & 1.791030 & 1.800700 & 1.803040 & 1.809010 & N/A \\
		Lower bound in \cite{Rudin01} & 1.748335 & 1.774098 & 1.792668 & 1.803471 & 1.808961 & 1.814327 & N/A \\
		This paper  & 1.749160 & 1.774240 & 1.793147 & 1.804701 & 1.814990 & 1.822499 & 1.829680 \\
		Improvement & 0.000825 & 0.000142 & 0.000479 & 0.001230 & 0.006029 & 0.008172 & N/A \\
		\hline
		$m$ & 12 & 13 & 14 & 15 & 16 & 17 & 18 \\
		\hline
		Lower bound in \cite{Rudin01} & 1.825277 & 1.829976 & 1.836791 & 1.839335 & 1.842994 & N/A & 1.845481 \\
		This paper  & 1.835064 & 1.838717 & 1.843699 & 1.848130 & 1.850929 & 1.852558 & 1.855025 \\
		Improvement & 0.009787 & 0.008741 & 0.006908 & 0.008795 & 0.007935 & N/A & 0.009544 \\
		\hline
		$m$ & 19 & 20 & 24  & 120 & 3600 & 8192 & arbitrary \\
		\hline
		Lower bound in \cite{Rudin01} & N/A & N/A & 1.856699  & 1.875000 & 1.880000 & N/A & 1.880000\\
		This paper  & 1.857616 & 1.859272 & 1.863644 & 1.883184 & 1.887095 & 1.887111 & 1.887111 \\
		Improvement & N/A & N/A & 0.006945  & 0.008184 & 0.007095 & N/A& 0.007111 \\
		\hline
	\end{tabular}
	\caption{Comparisons of lower bounds.}
\label{tab:comparison}
\end{table}

\textbf{Related Work:} Even before the concept of online problems was formally introduced, Graham \cite{Graham66} proposed the online algorithm List Scheduling, which was later proven to be optimal for $m=2$ and $m=3$ \cite{FKT89}. Surprisingly, despite continuous improvements in algorithms and lower bounds for both arbitrary and fixed numbers of machines, List Scheduling remains the sole known optimal algorithm for parallel machine scheduling to date.

For an arbitrary number of machines, the currently best algorithm achieves a competitive ratio of $1.9201$, a result proposed more than two decades ago \cite{FW00}. The most recent improvement to the lower bound also dates back over twenty years. Albers \cite{Albers99} showed that the lower bound is at least $1.852$. Gormley et al.~\cite{GRTW00} established a lower bound of $1.85358$
via computer search. Rudin~\cite{Rudin01} proved a lower bound of
$1.88$ in his dissertation. For a fixed small number of machines, lower bounds were established by Chen et al. \cite{CVW94}. 
Rudin improved these bounds, but only the lower bound of $\sqrt{3}$ for the $m=4$ case was formally published \cite{RC03}, with the proof later simplified by Braun et al. \cite{BCG25}. More recently, Janke reported a lower bound of $1.73246$ for $m=4$ \cite{Janke26}.
For a comprehensive overview on online parallel machine scheduling problems and algorithms, readers are referred to the survey \cite{PST04, TZ13}. 

\textit{Bin packing} is another important class of combinatorial optimization problems. In the one-dimensional bin packing problem, we are given a sequence of items, each with an associated size. The goal is to pack these items into the minimum number of bins, each having the same capacity. The \textit{First Fit Decreasing} (FFD) algorithm is a cornerstone for solving the bin packing problem \cite{Johnson73}. FFD first sorts items in non-increasing order of size and then assigns each item to the first open bin with sufficient remaining capacity. If no such bin exists, a new bin is opened. While research on FFD originated with Johnson's seminal work \cite{Johnson73}, it took over three decades for D\'{o}sa \cite{Dosa07} to fully characterize its worst-case performance.

The one-dimensional bin packing problem can be viewed as the dual of parallel machine scheduling. Bins correspond to machines, items to jobs, and item sizes to processing times. While parallel machine scheduling fixes the number of machines to minimize the makespan, bin packing fixes the makespan to minimize the number of machines utilized. This duality was recognized early in the literature and has been effectively leveraged to derive scheduling algorithms from bin packing algorithms \cite{CGJ78}.

The remainder of this paper is organized as follows. Section 2 details the construction of lower bounds for the general case. Section 3 focuses on the specific case of five machines. Finally, Section 4 provides some concluding remarks.

\section{Arbitrary Number of Machines}

In this section, we present a construction of instances for proving lower 
bounds for the problem on an arbitrary number of machines. We first 
introduce the general framework of the instances and explain why it can be 
used to establish lower bounds. We then describe a concrete procedure for 
finding such instances. Finally, we present the lower bounds obtained via 
our approach, together with the corresponding instances.

\subsection{General Framework}

Given a positive integer $N$, an integer $1\leq k < N$, 
and an $N$-tuple of real numbers $\mathbf{r}=(r_1,r_2,\dots,r_N)$, where
\begin{equation}\label{r7}
	1 = r_1 < r_2 < \cdots < r_N.
\end{equation}
Let
\begin{equation}\label{m6}
	\rho(\mathbf{r})=\min_{1\le i \le k}\frac{r_i}{\sum_{l=1}^{i} r_l},
\end{equation}
and for any $k+1\leq i\leq N$ and any $0<\gamma\leq \rho(\mathbf{r})$, define
\begin{equation}\label{m7}
	U_i(\gamma) = \frac{1}{\gamma}\sum_{l=1}^{i} r_l.
\end{equation}
Obviously, $\rho(\mathbf{r})\leq 1$ for any tuple $\mathbf{r}$ and $U_i(\gamma)\geq \sum_{l=1}^{i} r_l\geq  r_i$ for any $i$.

To prove the lower bound for $m$ machines, we construct the following 
instance. The jobs in the instance are partitioned into $N+1$ phases. 
Every phase except the last one consists of $m$ jobs, while the last phase contains only a single job with processing time $U_N(\gamma)$. In the first $k$ phases, all jobs have the same processing time. 
The processing time of jobs in Phase $i$ is $r_i$, $i = 1, \dots, k$. For any $k+1\leq i\leq N$, there are $Q+1$ types of jobs in Phase $i$, where 
$Q$ is a positive integer whose value is determined later in the 
implementation.
The processing time and the number of jobs of the $j$-th type in Phase $i$ are $p_{ij}$ and $n_{ij}$, respectively. We have $\sum_{j=1}^{Q+1} n_{ij} = m$, and the exact values of $n_{ij}$ will be determined later.
Note that for some $j$, we may have $n_{ij} = 0$, in which case jobs of this type do not actually exist. This is done only for notational convenience. 
For any $k+1\leq i\leq N$  and $1 \le j \le Q+1$, let
\begin{equation}\label{m11}
	p_{ij}=r_i+\frac{j-1}{Q}(U_i(\gamma) - r_i). 
\end{equation} 
Obviously, for any $k+1\leq i\leq N$,
\begin{equation}\label{m13}
	r_i = p_{i1} < p_{i2} < \dots < p_{iQ}<p_{i,Q+1}=U_i(\gamma).
\end{equation}

We now give a lemma that specifies when a good online algorithm has to assign the jobs in each phase to distinct machines. For any $i$ with $1\le i\le k$ and $j=1$, and for any $i$ with $k+1\le i\le N$ and $1\le j\le Q+1$, let $\mathcal{J}_{ij}$ be the instance consisting of jobs of the first $i-1$ phases and the first $j$ types of jobs of Phase $i$, and
\begin{equation}\label{m12}
	B_{ij}(\gamma) = \frac{ \sum_{l=1}^{i} r_l+p_{ij}}{1+\gamma}.
\end{equation}
By the definition of the instance, it is not difficult to see that 
\begin{equation}\label{r8}
C^*(\mathcal{J}_{i1})= \sum_{l=1}^{i} r_l, \quad 1\leq i\leq k.	
\end{equation}
For the remaining instances $\mathcal{J}_{ij}$, however, $C^*(\mathcal{J}_{ij})$
admits no simple expression in terms of the processing times. In this 
case, $B_{ij}(\gamma)$ serves as a good substitute for $C^*(\mathcal{J}_{ij})$.

\begin{lemma}\label{l1}
	Suppose that
	\begin{equation}\label{m4}
C^*(\mathcal{J}_{ij}) \le B_{ij}(\gamma), \quad \text{for all } k+1\leq i\leq N, \; 1 \le j \le Q+1 \text{ with } n_{ij}>0.
	\end{equation}
Any algorithm $A$ whose competitive ratio is less than $1+\gamma$, where 
$0 < \gamma \le \rho(\mathbf{r})$, must assign the jobs of the $i$-th 
phase to distinct machines for all $1 \le i \le N$.
\end{lemma}
\begin{proof}
	Assume that algorithm $A$ has a competitive ratio of less than $1+\gamma$. If  algorithm $A$ assigns at least two jobs of the first phase to the same machine, then $C^A \geq 2r_1$ and $\frac{C^A}{C^*} \geq \frac{2r_1}{r_1} = 2 \geq  1+\gamma$.  Therefore, jobs of the first phase must be processed on different machines.  
	Suppose that for $2\leq i\leq k$, the jobs in each of the first $i-1$ phases are scheduled on different machines. If the algorithm assigns at least two jobs of Phase $i$ to the same machine, then $C^A \geq  \sum_{l=1}^{i-1} r_l + 2r_{i}$. 
	Thus by \eqref{r8} and \eqref{r7},
	\begin{align*}
		\frac{C^A}{C^*} &\geq \frac{\sum_{l=1}^{i-1} r_l + 2r_{i}}{\sum_{l=1}^{i} r_l}  =1+\frac{r_{i}}{\sum_{l=1}^{i} r_l}\geq 1+\rho(\mathbf{r})\geq 1+\gamma.
	\end{align*}
	Therefore, jobs of Phase $i$ must be processed on different machines.
	By induction, the jobs in each of the first $k$ phases are scheduled on distinct machines.
	
	Suppose that for $k+1\leq i\leq N$, the jobs in each of the first $i-1$ phases are scheduled on different machines. If at least two jobs of the first type of Phase $i$ are processed on the same machine, then $C^A(\mathcal{J}_{i1}) \geq \sum_{l=1}^{i-1} r_l + 2p_{i1}$. By \eqref{m4} and \eqref{m13},
	$$\frac{C^A(\mathcal{J}_{i1})}{C^*(\mathcal{J}_{i1})} \ge \frac{\sum_{l=1}^{i-1} r_l + 2p_{i1}}{B_{i1}(\gamma)} = \frac{\sum_{l=1}^{i} r_l + p_{i1}}{B_{i1}(\gamma)} =1+\gamma.$$ 
	Therefore, jobs of the first type of Phase $i$ must be processed on different machines. Assume that the first $j-1$ types of jobs in Phase $i$ are all processed on different machines. If a job of the $j$-th type in Phase $i$ is processed on the same machine as another job of the same phase, then $C^A(\mathcal{J}_{ij}) \geq \sum_{l=1}^{i-1} r_l + p_{i1}+p_{ij}$. By \eqref{m4} and  \eqref{m13},
	$$\frac{C^A(\mathcal{J}_{ij})}{C^*(\mathcal{J}_{ij})} \ge \frac{\sum_{l=1}^{i-1} r_l + p_{i1}+p_{ij}}{B_{ij}(\gamma)} = \frac{\sum_{l=1}^{i} r_l + p_{ij}}{B_{ij}(\gamma)} =1+\gamma.$$ 
	Therefore, jobs of the first $j$ types of Phase $i$ must be processed on different machines. By induction, jobs of Phase $i$ are assigned to the different machines, and hence so are the jobs in each of the first $N$ phases.	
\end{proof}

Verifying the key condition  \eqref{m4} of the Lemma \ref{l1} requires computing the optimal makespan for quite a lot of instances, which is common in proving lower bounds.
However, computing the exact optimal makespan is computationally intractable. Therefore, an easily computable and relatively tight upper bound on the optimal makespan is sometimes used instead. We address this issue by considering the dual problem: minimizing the number of machines required for a given bound on makespan, rather than minimizing the makespan for a fixed number of machines. Specifically, the optimal makespan for scheduling a set of jobs on $m$ machines is at most $b$ if and only if all jobs can be scheduled on at most $m$ machines with a load of at most $b$ on each machine. This reformulation enables the use of algorithms from the bin-packing literature. 

The following procedure \texttt{FFD} is based on the First Fit Decreasing algorithm of the bin-packing problem \cite{Johnson73}. Given a set of jobs and a target makespan bound $b$, \texttt{FFD} first sorts the jobs in non-increasing order of their processing times, and then assigns the jobs one by one in this order. Each job is assigned to the first activated machine whose load, after the job is added, does not exceed $b$. If no such machine exists, a new machine is activated and the job is assigned to it. If \texttt{FFD} uses at most $m$ machines, it yields a feasible
schedule whose makespan does not exceed $b$, which implies that the
optimal makespan is at most $b$ as well. However, the converse does
not necessarily hold. If the optimal makespan is at most $b$,
\texttt{FFD} with bound $b$ may still use more than $m$ machines.
This is one of the limitations of the approach based on \texttt{FFD}.

\begin{algorithm}
	\caption*{\textbf{Procedure $\texttt{FFD}(\mathcal{J}_0,b)$}}
	\begin{algorithmic}[1]
\Statex \textbf{Input:} a job set $\mathcal{J}_0$ whose processing
times are $a_1, a_2, \dots, a_s$, and a bound $b$.
\Statex \textbf{Output:} the number of machines used, or $\infty$ if at least one job has processing time greater than $b$.
\State Reindex the jobs such that $a_1\ge a_2\ge \dots \ge a_s$.
		\If{$a_1>b$} \State \Return $\infty$ \EndIf
		\State $c\gets 1$; $\ell_1\gets a_1$
		\For{$j=2,\ldots,s$}
		\State $v\gets 1$
		\While{$v\le c$ \textbf{and} $\ell_v+a_j>b$}
		\State $v\gets v+1$
		\EndWhile
		\If{$v=c+1$} \State $c\gets c+1$; $\ell_c\gets 0$ \EndIf
		\State $\ell_v\gets \ell_v+a_j$
		\EndFor
		\State \Return $c$
	\end{algorithmic}
\end{algorithm}

Using Procedure \texttt{FFD}, we can determine the values of $n_{ij} $ such that \eqref{m4} is satisfied. Starting from the first type of the Phase $(k+1)$, the number of jobs of each type in each phase is determined sequentially.

Given a fixed value of $\gamma$, for each $k+1\leq i\leq N$ and $1\leq j\leq Q+1$, find a sufficiently good value $n_{ij}$ such that 
$C^*(\mathcal{J}_{ij}) \le B_{ij}(\gamma)$. This 
process can be implemented as follows. 
First, apply Procedure \texttt{FFD} to $\mathcal{J}_{i,j-1}$ under the makespan 
bound $B_{ij}(\gamma)$, where, for notational convenience, $\mathcal{J}_{k+1,0}$ is the set of jobs from 
the first $k$ phases, and $\mathcal{J}_{i,0} = \mathcal{J}_{i-1,Q+1}$ for $k+1 < i \le N$. While the 
procedure uses fewer than $m$ machines, we augment $\mathcal{J}_{i,j-1}$ 
with $d$ jobs of processing time $p_{ij}$, where $d$ is the number of 
currently unused machines, and run \texttt{FFD} again. Whenever the augmented job set still requires at 
most $m$ machines, the $d$ jobs are kept and the step is repeated. The 
process terminates as soon as \texttt{FFD} uses at least $m$ machines. Afterwards, further jobs are added one by one until more than $m$ machines are needed.
During the above process, the largest number of added jobs for which the number of used machines does not exceed $m$ is taken as the value of $n_{ij}$.

To ensure that the number of jobs at each phase is exactly $m$, we stop 
adding jobs as soon as the number reaches $m$ within a phase, and set the 
numbers of jobs of all subsequent types in that phase to $0$. Even so, 
it may still happen that $\sum_{j=1}^{Q+1} n_{ij} < m$ for some 
$k+1\le i\le N$, in which case the construction fails.
If no failure occurs 
up to phase $N$ but $\texttt{FFD}(\mathcal{J}, U_N(\gamma)) > m$, the construction fails 
as well. Otherwise, the construction succeeds. The following theorem establishes the theoretical significance for such successfully constructed instances.

\begin{theorem}\label{thm2}
	If for given $m$ and  $\gamma$, the construction is successful, then there does not exist an algorithm with competitive ratio smaller than $1+\gamma$.
\end{theorem}
\begin{proof}
	Assume that algorithm $A$ has a competitive ratio of less than $1+\gamma$. Since the construction does not fail at any Phase $i$, each phase has exactly $m$ jobs. Moreover, for all $k+1\leq i\leq N$ and $1 \le j \le Q+1$ with $n_{ij}>0$, $\texttt{FFD}(\mathcal{J}_{ij}, B_{ij}(\gamma)) \le m$ by the choice of $n_{ij}$, and hence $C^*(\mathcal{J}_{ij}) \le B_{ij}(\gamma)$. By Lemma \ref{l1}, the jobs of each of the first $N$ phases are processed on different machines. Thus, after the last job arrives, $C^A(\mathcal{J}) \ge \sum_{i=1}^{N} r_i + U_N(\gamma)$, regardless of which machine it is processed on. And since the construction does not fail, $\texttt{FFD}(\mathcal{J}, U_N(\gamma)) \le m$, hence $C^*(\mathcal{J}) \le U_N(\gamma)$. Therefore, by \eqref{m7}, we have
	$$\frac{C^A(\mathcal{J})}{C^*(\mathcal{J})} \ge \frac{\sum_{i=1}^{N} r_i + U_N(\gamma)}{U_N(\gamma)} = 1+\gamma,$$
	which contradicts the assumption that the competitive ratio of $A$ is 
	less than $1+\gamma$.
\end{proof}

\subsection{Example of Construction}

In this subsection, we illustrate the complete process of the instance construction using a toy example. Suppose that $m = 6$, $N = 3$, $k = 1$, and $\mathbf{r}=(1, 2, 3)$. We fix $\gamma=\frac{2}{3}$ and choose $Q = 3$. Then, by \eqref{m7}, we have $U_2(\gamma)=\frac{9}{2}$ and $U_3(\gamma)=9$. The corresponding values of $p_{ij}$ and $B_{ij}(\gamma)$ for $i=2,3$ and $1\leq j\leq 4$, obtained by \eqref{m11} and \eqref{m12}, are shown in Table~\ref{ex1}. 

\begin{table}[htbp]
	\centering
	\renewcommand{\arraystretch}{1.25}
	\begin{tabular}{|c|c|c|c|c|c|c|c|c|c|c|}
		\hline
		& \multicolumn{2}{c|}{Phase $1$} & \multicolumn{3}{c|}{Phase $2$} & \multicolumn{3}{c|}{Phase $3$} & \multicolumn{2}{c|}{Phase $4$} \\
		\hline
		$j$ & $r_1$ & number & $p_{ij}$ & $B_{ij}$ & $n_{ij}$ & $p_{ij}$ & $B_{ij}$ & $n_{ij}$ & $U_3(\gamma)$ & number \\
		\hline
		1 & \multirow{4}{*}{$1$} & \multirow{4}{*}{$6$} & $2$            & $3$           & $6$ & $3$ & $\frac{27}{5}$ & $4$ & \multirow{4}{*}{$9$} & \multirow{4}{*}{$1$} \\
		\cline{1-1}\cline{4-9}
		2 &  &  & $\frac{17}{6}$ & $\frac{7}{2}$ & $0$ & $5$ & $\frac{33}{5}$ & $1$ &  &  \\
		\cline{1-1}\cline{4-9}
		3 &  &  & $\frac{11}{3}$ & $4$           & $0$ & $7$ & $\frac{39}{5}$ & $1$ &  &  \\
		\cline{1-1}\cline{4-9}
		4 &  &  & $\frac{9}{2}$  & $\frac{9}{2}$ & $0$ & $9$ & $9$ & $0$ &  &  \\
		\hline
	\end{tabular}
	\caption{Overview of the instance} \label{ex1}
\end{table}
	
According to the structure of the instance, Phase $1$ contains $6$ jobs with processing time $r_1=1$, and Phase $4$ consists of a single job with processing time $U_3(\gamma)=9$. Next, we determine the number of jobs of each type in Phases $2$ and $3$.	

Note that $\texttt{FFD}(\mathcal{J}_{2,0},3)=2$, and the jobs of the first type in Phase~$2$ have a processing time of $2$. Adding $4$ more jobs increases the number of machines used by \texttt{FFD} to $5$, while adding one additional job raises this number to $6$. Even with the inclusion of yet another such job, \texttt{FFD} still requires $6$ machines. Since Phase~$2$ already contains $6$ jobs at this point, we have $n_{21}=6$ and $n_{22}=n_{23}=n_{24}=0$. Thus, the numbers of jobs of each type in Phase~$2$ are fully determined.

Similarly, $\texttt{FFD}(\mathcal{J}_{3,0},\frac{27}{5})=4$, and the jobs of the first type in Phase~$3$ have a processing time of $3$. Adding $2$ more jobs increases the number of machines used by \texttt{FFD} to $5$, while adding one additional job raises this number to $6$. However, upon adding one additional job, \texttt{FFD} requires $7$ machines. Hence, $n_{31}=4$. Note that $\texttt{FFD}(\mathcal{J}_{3,1},\frac{33}{5})=5$, and the jobs of the second type in Phase~$3$ have a processing time of $5$. Adding one more job increases the machine count to $6$. However, upon adding one additional job, \texttt{FFD} requires $7$ machines. Hence, $n_{32}=1$. Similarly, $\texttt{FFD}(\mathcal{J}_{3,2},\frac{39}{5})=5$, and the jobs of the third type in Phase~$3$ have a processing time of $7$. Adding one more job increases the machine count to $6$. However, upon adding one additional job, \texttt{FFD} requires $7$ machines. Hence, $n_{33}=1$. Since Phase~$3$ already contains $6$ jobs, $n_{34}=0$. Thus, the number of jobs for each type in Phase~$3$ is completely determined.

Finally, after the last job is added to the job set, $\texttt{FFD}(\mathcal{J},9)=6$. Therefore, the construction is successful. 

Table~\ref{ex2} reports the schedule of the job set 
produced by \texttt{FFD} during this process. The first column displays the initial job set used to determine the number of jobs for each type in each phase, and the second column indicates the number of such jobs incrementally added to the previous job set. The last column reports the number of machines used by \texttt{FFD}. If this number reaches $7$, the job is not actually added to the job set.

\begin{table}[htbp]
	\centering
	\begin{tabular}{|c|c|c|c|c|c|c|c|c|c|}
		\hline
		  &   & $M_1$ & $M_2$ & $M_3$ & $M_4$ & $M_5$ & $M_6$ & $M_7$  &  \\
		\hline
		$\mathcal{J}_{2,0}$	& $0$ & $\{1,1,1\}$ & $\{1,1,1\}$ &  &  &  &  &  & $2$ \\
		\hline
		&   $4$ & $\{2,1\}$ & $\{2,1\}$ & $\{2,1\}$ & $\{2,1\}$ & $\{1,1\}$ & & & $5$ \\
		\hline
		& $1$ & $\{2,1\}$ & $\{2,1\}$ & $\{2,1\}$  & $\{2,1\}$ & $\{2,1\}$ & $\{1\}$ &	& $6$ \\
		\hline
		& $1$ & $\{2,1\}$ & $\{2,1\}$ & $\{2,1\}$  & $\{2,1\}$ & $\{2,1\}$ & $\{2,1\}$ & & $6$ \\
		\hline
		$\mathcal{J}_{3,0}$	& $0$ & $\{2,2,1\}$ & $\{2,2,1\}$ & $\{2,2,1\}$ & $\{1,1,1\}$ &  &  &  & $4$ \\
		\hline
		&   $2$ & $\{3,2\}$ & $\{3,2\}$ & $\{2,2,1\}$ & $\{2,2,1\}$ & $\{1,1,1,1\}$ &   & & $5$ \\
		\hline
		& $1$ & $\{3,2\}$ & $\{3,2\}$ & $\{3,2\}$ & $\{2,2,1\}$ & $\{2,1,1,1\}$ &  $\{1,1\}$ & & $6$ \\
		\hline
		& $1$ & $\{3,2\}$ & $\{3,2\}$ & $\{3,2\}$ & $\{3,2\}$ & $\{2,2,1\}$ &  $\{1,1,1,1,1\}$ & & $6$ \\
		\hline
		& $1$ & $\{3,2\}$ & $\{3,2\}$ & $\{3,2\}$ & $\{3,2\}$ & $\{3,2\}$ &  $\{2,1,1,1,1\}$ & $\{1,1\}$ & $7$ \\
		\hline
		$\mathcal{J}_{3,1}$	& $0$ & $\{3,3\}$ & $\{3,3\}$ & $\{2,2,2\}$ & $\{2,2,2\}$ & $\{1,1,1,1,1,1\}$ &  &  & $5$ \\
		\hline
		&   $1$ & $\{5,1\}$ & $\{3,3\}$ & $\{3,3\}$ & $\{2,2,2\}$ & $\{2,2,2\}$ &  $\{1,1,1,1,1\}$ & & $6$ \\
		\hline
		& $1$ & $\{5,1\}$ & $\{5,1\}$ & $\{3,3\}$ & $\{3,3\}$ & $\{2,2,2\}$ & $\{2,2,2\}$ &  $\{1,1,1,1\}$  & $7$ \\
		\hline
		$\mathcal{J}_{3,2}$	& $0$ & $\{5,2\}$ & $\{3,3,1\}$ & $\{3,3,1\}$ & $\{2,2,2,1\}$ & $\{2,2,1,1,1\}$ &  &  & $5$ \\
		\hline
		&   $1$ & $\{7\}$ & $\{5,2\}$ & $\{3,3,1\}$ & $\{3,3,1\}$ & $\{2,2,2,1\}$ & $\{2,2,1,1,1\}$ & &  $6$ \\
		\hline
		$\mathcal{J}$ & & $\{9\}$	& $\{7,2\}$ & $\{5,3,1\}$ & $\{3,3,3\}$ & $\{2,2,2,2,1\}$ & $\{2,1,1,1,1\}$ &  &   $6$ \\
		\hline
	\end{tabular}
	\caption{Schedules produced by \texttt{FFD}.}\label{ex2}
\end{table}

\subsection{Procedure \texttt{Construct}}

By Theorem \ref{thm2}, it suffices to successfully construct such an instance 
to establish the lower bound. The detailed steps are outlined in Procedure \texttt{Construct}.
Given the number of machines $m$, the selected values of $N$
and $k$, the tuple $\mathbf{r}$, and a candidate lower bound $\gamma$, the
procedure returns \textsc{true} if an instance can be constructed successfully, and
\textsc{false} otherwise. We remark that since \texttt{FFD} only provides an estimate of the optimum of an instance rather than computing it exactly, a return value 
of \textsc{false} does not necessarily imply that the lower bound is 
smaller than $\gamma$. Consequently, the fact that $\texttt{Construct}(m, N, k, \mathbf{r}, \gamma)$ returns \textsc{false} does not guarantee that $\texttt{Construct}(m, N, k, \mathbf{r}, \gamma')$ will also return \textsc{false} for any $\gamma'>\gamma$.

Throughout the procedure, we write $\{a\}^{i}$ for the multiset 
containing $i$ copies of $a$, and $P_1 \uplus P_2$ for the multiset 
union of two multisets $P_1$ and $P_2$. For each Phase $i$, $P_{ij}$ 
denotes the multiset of the processing times of the jobs in 
$\mathcal{J}_{ij}$, and $P_{i0}$ denotes the multiset of the processing 
times of the jobs in the first $i-1$ phases. Moreover, $C_i$ counts the 
number of jobs in Phase $i$ that have been included in the instance.

\begin{algorithm}[!htbp]
	\caption*{\textbf{Procedure $\texttt{Construct}(m, N, k, \mathbf{r}, \gamma)$}}
	\begin{algorithmic}[1]
\Statex \textbf{Input:} the number of machines $m$, integers $N$ and $k$,
an $N$-tuple $\mathbf{r}=(r_1,\dots,r_N)$ with $1=r_1 < r_2 < \cdots < r_N$, and a bound $\gamma$.
\Statex \textbf{Output:} \textsc{true} if the construction succeeds, and
\textsc{false} otherwise.
		\If{$\gamma>\rho(\mathbf{r})$}
		\State \Return \textsc{false}
		\EndIf
		\State $P_{k+1,0}\gets \biguplus_{l=1}^k\{r_l\}^m$.
		\For{$i=k+1,\ldots,N$}
		\State $U_i(\gamma)\gets \frac{1}{\gamma}\sum_{l=1}^{i} r_l$
		\State $C_i\gets 0$; set $n_{ij}\gets 0$ for all $j=1,\ldots,Q+1$.
		\For{$j=1,\ldots,Q+1$}
		\State $P_{ij}\gets P_{i,j-1}$.
		\If{$C_i=m$}
		\State set $P_{i\ell}\gets P_{ij}$ for $\ell=j+1,\ldots,Q+1$; \textbf{break}
		\EndIf
		\State $p_{ij}\gets r_i+\frac{j-1}{Q}(U_i(\gamma) - r_i)$, $B_{ij}(\gamma) \gets \frac{ \sum_{l=1}^{i} r_l+p_{ij}}{1+\gamma}$
\If {$\texttt{FFD}(P_{ij},B_{ij}(\gamma))=\infty$}
\State \Return \textsc{false} \EndIf
		\While{$C_i<m$}
		\State $d\gets\min\{m-C_i,\max\{0,m-\texttt{FFD}(P_{ij},B_{ij}(\gamma))\}\}$.
\If{$d>0$ \textbf{and}  $\texttt{FFD}(P_{ij}\uplus\{p_{ij}\}^d,B_{ij}(\gamma))\leq  m$} 
		\State $P_{ij}\gets P_{ij} \uplus \{p_{ij}\}^d$;
$n_{ij}\gets n_{ij}+d$; $C_i\gets C_i+d$.
\Else \State \textbf{break} \EndIf
		\EndWhile
		\While{$C_i<m$ \textbf{and} $\texttt{FFD}(P_{ij}\uplus\{p_{ij}\},B_{ij}(\gamma))\le m$}
		\State $P_{ij}\gets P_{ij} \uplus \{p_{ij}\}$; $n_{ij}\gets n_{ij}+1$; $C_i\gets C_i+1$.
		\EndWhile
		\EndFor
		\If{$C_i<m$} \State \Return \textsc{false} \EndIf
        \State  $P_{i+1,0}\gets P_{i,Q+1}$ 
		\EndFor
		\If{$\texttt{FFD}(P_{N,Q+1}\uplus \{U_N(\gamma)\},U_N(\gamma))>m$}
		\State \Return \textsc{false}
		\EndIf
		\State $P \gets P_{N,Q+1}\uplus \{U_N(\gamma)\}$.
		\State \Return \textsc{true}
	\end{algorithmic}
\end{algorithm}

The Procedure \texttt{Construct} not only determines whether the desired instance can be constructed, but also yields the multiset $P$, upon success, which comprises the processing times of all jobs in the instance. 

\subsection{Binary Search}

The success of the construction depends not only on the choices of $N$, $k$, 
and $\mathbf{r}$, but also on the choice of $\gamma$. To date, no criterion 
is available for identifying combinations of these parameters that 
necessarily lead to success. Therefore, we adopt the binary search for 
$\gamma$ 
to find a satisfactory value of 
$\gamma^*$ for which the construction succeeds.

We initialize the search by setting $\gamma=\rho(\mathbf{r})$. If the construction
succeeds, then $\rho(\mathbf{r})$ is the desired value. Otherwise, we set
$\gamma_{LB}=\gamma_0$ and $\gamma_{UB}=\rho(\mathbf{r})$ as lower and upper bounds on
the search interval, where $\gamma_0$ can be set to a known lower bound or simply some value that we hope the resulting lower bound will exceed.  In each iteration, we set
$\gamma=\frac{1}{2}(\gamma_{LB}+\gamma_{UB})$ and test whether the
construction succeeds. If it does, we set $\gamma_{LB}=\gamma$. Otherwise, we set $\gamma_{UB}=\gamma$. The procedure terminates
once the difference $\gamma_{UB}-\gamma_{LB}$ falls below a prescribed
threshold $\delta$.

The complete procedure is presented in Procedure \texttt{BinarySearch}. Given the number of machines $m$, the selected values of $N$ and $k$, 
the tuple $\mathbf{r}$, and a known lower bound $\gamma_0$, the 
procedure returns the largest value $\gamma^{*}>\gamma_0$ that our 
approach can certify as an improved lower bound, or $\bot$ indicating 
that no such value can be found. Nevertheless, since a return value of 
\textsc{false} from \texttt{Construct} does not imply that the 
corresponding lower bound does not exist, the returned $\gamma^{*}$ 
need not be the best possible lower bound for the given parameters 
$N$, $k$, and $\mathbf{r}$. Moreover, returning $\bot$ does not rule 
out the possibility that the lower bound can be improved.

\begin{algorithm}[H]
\caption*{\textbf{Procedure $\texttt{BinarySearch}(m, N, k, \mathbf{r}, \gamma_0)$}}	
	\begin{algorithmic}[1]
\Statex \textbf{Input:} the number of machines $m$, integers $N$ and $k$, an $N$-tuple $\mathbf{r}=(r_1,\dots,r_N)$ with $1=r_1 < r_2 < \cdots < r_N$, and an initial lower bound $\gamma_0$.
\Statex \textbf{Output:} a lower bound $\gamma^{*}>\gamma_0$ or $\rho(\mathbf{r})$ certified by the
construction, or $\bot$ if no value greater than $\gamma_0$ succeeds.
		\If{$\texttt{Construct}(m,N,k,\mathbf{r},\gamma_0)=\textsc{false}$}
        \State \Return $\bot$
        \EndIf
		\If{$\texttt{Construct}(m,N,k,\mathbf{r},\rho(\mathbf{r}))=\textsc{true}$}
		\State \Return $\rho(\mathbf{r})$
		\EndIf
		\State $\gamma_{LB}\gets \gamma_0$; $\gamma_{UB}\gets \rho(\mathbf{r})$; $\gamma^{*}\gets\bot$
		\While{$\gamma_{UB}-\gamma_{LB}>\delta$}
		\State $\gamma\gets\frac{1}{2}(\gamma_{LB}+\gamma_{UB})$
		\If{$\texttt{Construct}(m,N,k,\mathbf{r},\gamma)=\textsc{true}$}
		\State $\gamma_{LB}\gets\gamma$; $\gamma^{*}\gets\gamma$
		\Else
		\State $\gamma_{UB}\gets\gamma$
		\EndIf
		\EndWhile
		\State \Return $\gamma^{*}$
	\end{algorithmic}
\end{algorithm}

\subsection{Initial Lower bound and Local Search}

In the previous subsection, we presented the general framework of our construction, leaving many parameters unspecified. In this subsection, we make the values of these parameters explicit, with emphasis on the construction of the tuple $\mathbf{r}$. We start by presenting the exact values of an initial choice of $\mathbf{r}$ for fixed $N$ and $k$. If it does not lead to a satisfactory lower bound, local search is applied to obtain better tuples.

Let $\mu \in (0,1)$ be a parameter whose value is to be specified later.
Let
\begin{equation}\label{m8}
	r_i = \frac{\mu}{(1-\mu)^{i-1}}r_1, \quad 2\leq i\leq k,
\end{equation}
and
\begin{equation}\label{m9}
	r_i =  \frac{ 2\mu r_{i-1}-\sum_{l=1}^{i-1} r_l}{1-\mu}, \quad k+1\leq i\leq N.	
\end{equation}
For $2\leq i\leq k$, by \eqref{m8},
\begin{align*}
	\frac{r_{i}}{\sum_{l=1}^{i} r_l}=\frac{\frac{\mu}{(1-\mu)^{i-1}}r_1}{\left( 1+\sum_{l=2}^{i} \frac{\mu}{(1-\mu)^{l-1}}\right) r_1}=\frac{\frac{\mu}{(1-\mu)^{i-1}}}{1+\frac{\frac{\mu}{1-\mu}\left( 1-\frac{1}{(1-\mu)^{i-1}} \right)  }{1-\frac{1}{1-\mu} }}=\mu.
\end{align*}
Thus, $\rho(\mathbf{r})=\mu$ for $1<k<N$, and $\rho(\mathbf{r})=1$ for $k=1$. 
Setting $\gamma=\mu$, the processing time of the last job becomes
$$
U_N(\mu) = \frac{1}{\mu}\sum_{l=1}^{N} r_l = \frac{1}{\mu}\left(2\mu r_{N-1}-(1-\mu)r_{N}+r_{N} \right) = r_N+2r_{N-1}
$$
by \eqref{m9}. Consequently, $C^*(\mathcal{J})\geq r_N+2r_{N-1}$. 
On the other hand, there are $m$ jobs in phase $N$ with processing times of at least $r_N$, and the last job has a processing time of $U_N(\gamma)>r_N$. Thus, $C^*(\mathcal{J})\geq 2r_N$. 
Therefore, following a standard technique in competitive analysis, we choose $\mu$ such that $r_N = 2r_{N-1}$, which helps make the machine loads in the optimal schedule roughly equal.
Under this condition, $U_N(\gamma) < U_N(\mu) = 2r_N \leq C^*(\mathcal{J})$ whenever $\gamma>\mu$. Since the construction succeeds only if $\texttt{FFD}(\mathcal{J}, U_N(\gamma)) \leq m$, 
this is possible only when $\gamma \leq \mu$, even in the case $k=1$.

We choose $N$ and $k$ such that the corresponding values of $1+\mu$ are candidates for the lower bound. The tuple of $\mathbf{r}$ can be obtained by \eqref{m8} and \eqref{m9}. 
Given a value $\gamma_0$, the procedure $\texttt{BinarySearch}(m, N, k, 
\mathbf{r}, \gamma_0)$ attempts to find a lower bound strictly better than 
$\gamma_0$. If the procedure fails to find such a lower bound, one can change 
the values of $N$ and $k$ and repeat the above procedure with the new 
parameters. If the procedure finds such a lower bound but the result is unsatisfactory, one can either change the values of $N$ and $k$ and repeat the above process, or keep them unchanged and apply the local search approach described below, which attempts to obtain a better lower bound by perturbing $\mathbf{r}$.


For given values of $N$ and $k$, let $\mathbf{r}^{(0)}=(r_1^{(0)}, \dots, r_N^{(0)})$ and $\gamma^{(0)}$ denote the initial tuple obtained by \eqref{m8} and \eqref{m9} and the corresponding lower bound, respectively, serving as the starting point of the current run. However, the initial tuple and lower bound are not limited to these values. Alternatively, the tuple and lower bound obtained from a previous run of the procedure can serve as the starting point for the next run.

Suppose that the parameters obtained in iteration $t$ are $\mathbf{r}^{(t)}$ and $\gamma^{(t)}$. Choose a step size $h$, take $\boldsymbol{\varepsilon}=(0, \varepsilon_2, \dots, \varepsilon_N)$, where $|\varepsilon_i| \le h$ for $i=2, \dots, N$, and let $\mathbf{r}=(r_1, r_2, \dots, r_N)$, where $r_i = r_i^{(t-1)}(1+\varepsilon_i)$ for $i=2, \dots, N$. If $\mathbf{r}$ does not satisfy \eqref{r7}, the tuple is infeasible. In this case, $\mathbf{r}$ is rejected, thereby requiring $\boldsymbol{\varepsilon}$ to be regenerated, and the procedure proceeds to the next iteration.
If $\mathbf{r}$ is feasible, apply $\texttt{BinarySearch}(m, N, k, 
\mathbf{r}, \gamma^{(t-1)})$. If it returns $\bot$, indicating that no lower bound better than $\gamma^{(t-1)}$ can be found, or even that no valid lower bound can be obtained at all, the tuple $\mathbf{r}$ is rejected. This necessitates the regeneration of $\boldsymbol{\varepsilon}$, and the procedure advances to the next iteration. Otherwise, set $\mathbf{r}^{(t)}=\mathbf{r}$ 
and $\gamma^{(t)}=\gamma$, and proceed to the next iteration.
The iteration terminates after $T$ steps. If no satisfactory lower 
bound has been found by that point, $N$ and $k$ must be reselected.
 

The procedure \texttt{LocalSearch} is given below. The step size $h$ 
and the number of iterations $T$ involved in the procedure can be 
chosen according to practical needs. Each $\epsilon_i$ is drawn from a uniform distribution on $[-h, h]$, so that a different tuple can be obtained through the random perturbation.
 Given the 
output $\gamma^{**}$ and $\mathbf{r}^{**}$ of procedure
$\texttt{LocalSearch}(m, N, k, \mathbf{r}^{(0)},\gamma^{(0)})$, running procedure
$\texttt{Construct}(m, N, k, \mathbf{r}^{**}, \gamma^{**})$ yields the 
corresponding instance.

\begin{algorithm}[H]
	\caption*{\textbf{Procedure $\texttt{LocalSearch}(m,N,k,\mathbf{r}^{(0)},\gamma^{(0)})$}}
	\begin{algorithmic}[1]
		\Statex \textbf{Input:} the number of machines $m$, integers $N$ and $k$,
		an initial lower bound $\gamma^{(0)}$ and the corresponding	$N$-tuple $\mathbf{r}^{(0)}=(r_1^{(0)},\dots,r_N^{(0)})$ with $1=r_1^{(0)} < r_2^{(0)} < \cdots < r_N^{(0)}$.
		\Statex \textbf{Output:} a lower-bound $\gamma^{**}$ and the
		corresponding tuple $\mathbf{r}^{**}$.
		\For{$t=1,\dots,T$}
		\State draw $\epsilon_i\sim\mathrm{Uni}[-h,h]$
		for all $i\in\{2,\dots,N\}$
		\State set $ r_1\gets 1$ and
		$ r_i\gets r_i^{(t-1)}(1+\epsilon_i)$
		for $2\le i\le N$
		\If{$ r_1< r_2<\dots< r_N$
			does not hold}
		\State $\mathbf{r}^{(t)}\gets\mathbf{r}^{(t-1)}$;
		$\gamma^{(t)}\gets\gamma^{(t-1)}$
		\State \textbf{continue}
		\EndIf
		\State $\mathbf{r}\gets
		( r_1, r_2,\dots, r_N)$

		\State $\gamma\gets
\texttt{BinarySearch}(m,N,k,\mathbf{r},\gamma^{(t-1)})$
\If{$\gamma\ne\bot$}
\State $\mathbf{r}^{(t)}\gets\mathbf{r}$;
$\gamma^{(t)}\gets\gamma$
\Else  
\State $\mathbf{r}^{(t)}\gets\mathbf{r}^{(t-1)}$;
$\gamma^{(t)}\gets\gamma^{(t-1)}$
\EndIf

		\EndFor
		
		\State $\mathbf{r}^{**}\gets\mathbf{r}^{(T)}$;
		$\gamma^{**}\gets\gamma^{(T)}$
		\State \Return $\gamma^{**}$ and $\mathbf{r}^{**}$
	\end{algorithmic}
\end{algorithm}

\subsection{Computational Results}

In this subsection, we summarize the lower bounds obtained by implementing 
the methods described above. We first explain the principles behind the 
choice of the parameters used in the program. These choices are mostly 
empirical. Larger values of $Q$, $h$ and $T$ increase the chance of obtaining an instance that yields a larger lower bound, though this is not guaranteed, while at the same time making the computation more time-consuming. The remaining 
parameter $\delta$ is fixed at $10^{-8}$ to guarantee the desired accuracy.

For each fixed $m$, we started with a larger $h$ and smaller values of $Q$ and $T$, and then gradually increased $Q$ and $T$ or decreased $h$, using the parameter values drawn from $Q\in\{100,1000,2048,10000\}$,
$$h\in\{0.3,0.1,0.03,0.01,0.003,0.001,0.0003,0.0001,0.00003,0.00001,0.000003\},$$
and $T\in\{10,20,30,100,200,500,1000,1500,2000,5000\}$.

All computations were performed on a laptop computer equipped with an AMD Ryzen 7 8745H CPU (16 logical processors, 3.8 GHz base frequency) and 24 GB of RAM, running Windows 11 Home. The algorithms were implemented in Python 3.13.2. Each run took between $10$ seconds and $3$ hours. For instance, with $m=6$, $Q=1000$, $h\in\{0.3,0.1,0.03,0.01,0.003,0.001,0.0003,0.0001\}$, and $T=1500$, individual runs took about $10$ to $50$ seconds, and the eight runs took about $200$ seconds in total. For $m=2048$, $Q\in\{100,1000,2048\}$, $h\in\{0.01,0.001,0.0001,0.00003,0.00001\}$, and $T=10$, the $15$ parameter combinations required an estimated $2$ to $3$ hours in total, corresponding to approximately $10$ minutes per run on average.

Table \ref{tab:lower-bounds} summarizes the lower bounds obtained by our approach, with the corresponding instances provided in the supplementary material \cite{XT26}. It can be observed from the table that the initial lower bound is already quite 
good when $m$ is relatively large, while the local search brings a considerable 
improvement, especially when $m$ is relatively small.

\begin{table}[htbp]
	\centering
	\begin{tabular}{rrrcccc rrrcccc}
		\hline
		$m$ & $N$ & $k$ & $\mu$ & $1+\gamma^*$ & $1+\gamma^{**}$
		& & $m$ & $N$ & $k$ & $\mu$ & $1+\gamma^*$ & $1+\gamma^{**}$ \\
		\hline
6   & 8  & 1 & 0.876990 & 1.744366 & 1.774240 & & 27  & 13 & 3 & 0.883055 & 1.858084 & 1.866473 \\
7   & 7  & 1 & 0.873729 & 1.768565 & 1.793147 & & 28  & 12 & 3 & 0.882016 & 1.858263 & 1.867420 \\
8   & 8  & 2 & 0.876406 & 1.784615 & 1.804701 & & 29  & 10 & 2 & 0.880669 & 1.861721 & 1.868283 \\
9   & 7  & 1 & 0.873729 & 1.795486 & 1.814990 & & 40  & 8  & 1 & 0.876990 & 1.870433 & 1.872611 \\
10  & 7  & 1 & 0.873729 & 1.802418 & 1.822499 & & 50  & 8  & 1 & 0.876990 & 1.873609 & 1.874874 \\
11  & 10 & 2 & 0.880669 & 1.807120 & 1.829680 & & 54  & 9  & 2 & 0.878874 & 1.874902 & 1.876409 \\
12  & 12 & 1 & 0.883237 & 1.822577 & 1.835064 & & 68  & 10 & 1 & 0.880976 & 1.877163 & 1.879442 \\
13  & 7  & 1 & 0.873729 & 1.829957 & 1.838717 & & 72  & 10 & 2 & 0.880669 & 1.878142 & 1.879446 \\
14  & 7  & 2 & 0.872871 & 1.829689 & 1.843699 & & 92  & 11 & 2 & 0.882016 & 1.880028 & 1.881068 \\
15  & 7  & 1 & 0.873729 & 1.832254 & 1.848130 & & 99  & 12 & 2 & 0.883055 & 1.880354 & 1.881378 \\
16  & 9  & 2 & 0.878874 & 1.837561 & 1.850929 & & 108 & 12 & 1 & 0.883237 & 1.881112 & 1.882274 \\
17  & 10 & 2 & 0.880669 & 1.843359 & 1.852558 & & 112 & 13 & 2 & 0.883873 & 1.881179 & 1.882715 \\
18  & 14 & 1 & 0.884647 & 1.844327 & 1.855025 & & 120 & 14 & 2 & 0.884530 & 1.881492 & 1.883184 \\
19  & 11 & 3 & 0.880669 & 1.849440 & 1.857616 & & 128 & 14 & 2 & 0.884530 & 1.881786 & 1.883205 \\
20  & 12 & 3 & 0.882016 & 1.850440 & 1.859272 & & 256 & 16 & 1 & 0.885586 & 1.884353 & 1.884730 \\
21  & 13 & 3 & 0.883055 & 1.851633 & 1.860774 & & 512 & 17 & 1 & 0.885942 & 1.885906 & 1.885915 \\
22  & 13 & 3 & 0.883055 & 1.853224 & 1.862341 & & 1024& 20 & 2 & 0.886779 & 1.886274 & 1.886441 \\
23  & 12 & 3 & 0.882016 & 1.855232 & 1.863082 & & 2048& 24 & 1 & 0.887357 & 1.884898 & 1.886851 \\
24  & 11 & 2 & 0.882016 & 1.856041 & 1.863644 & & 3600& 21 & 1 & 0.886913 & 1.878676 & 1.887095 \\
25  & 12 & 3 & 0.882016 & 1.856699 & 1.864555 & & 4096& 21 & 1 & 0.886913 & 1.878676 & 1.887095 \\
26  & 12 & 3 & 0.882016 & 1.857110 & 1.865408 & & 8192& 21 & 1 & 0.886913 & 1.878676 & 1.887111 \\
		\hline
	\end{tabular}
	\caption{Lower bounds for different numbers of machines.}
\label{tab:lower-bounds}
\end{table}

Table \ref{tab:local-search} illustrates the first several steps of the local search, starting from the initial lower bound for the case with $m=6$, $N=8$, and $k = 1$. The parameter $h$ is set to $0.3$. The tuples obtained at each step are presented, and each step is categorized as either "Accept" or "Reject". In "Accept" steps, a better lower bound is obtained. 

\begin{table}[htbp]
	\centering
	\small
	\begin{tabular}{rrrrrrrrrcr}
		\hline
		Step & $r_2$ & $r_3$ & $r_4$ & $r_5$ & $r_6$ & $r_7$ & $r_8$ & $\gamma$ & Decision & $\gamma^{*}$ \\
		\hline
		0  & 6.129 & 29.440 & 122.493 & 453.522 & 1486.736 & 4132.853 & 8265.706 & --        & Initial & 0.744366 \\
		1  & 4.414 & 28.401 & 139.109 & 319.146 & 1373.938 & 3681.216 & 10268.889 & 0.770099 & Accept  & 0.770099 \\
		2  & 4.241 & 34.514 & 113.191 & 238.785 & 1249.996 & 2665.059 & 7972.625 & --        & Reject  & 0.770099 \\
		3  & 3.119 & 23.855 & 145.890 & 317.966 & 1744.028 & 3796.903 & 11167.705 & --       & Reject  & 0.770099 \\
		4  & 5.359 & 27.397 & 129.783 & 378.602 & 1197.137 & 3599.893 & 9883.748 & --        & Reject  & 0.770099 \\
		5  & 3.991 & 32.940 & 126.130 & 268.878 & 1092.924 & 4063.684 & 10272.140 & 0.772073 & Accept  & 0.772073 \\
		6  & 3.375 & 33.495 & 108.270 & 324.202 & 1004.438 & 5000.522 & 11899.926 & --       & Reject  & 0.772073 \\
		7  & 4.983 & 28.440 & 126.244 & 293.293 & 1057.580 & 4456.927 & 10968.492 & --       & Reject  & 0.772073 \\
		8  & 2.891 & 29.571 & 97.559 & 304.897 & 922.656 & 4069.974 & 9976.267 & --        & Reject  & 0.772073 \\
		9  & 3.204 & 31.632 & 109.279 & 251.674 & 1319.090 & 5221.449 & 12749.013 & --       & Reject  & 0.772073 \\
		10 & 3.765 & 36.024 & 103.060 & 341.795 & 1327.113 & 3049.875 & 8936.257 & --        & Reject  & 0.772073 \\
		11 & 3.940 & 38.542 & 135.746 & 289.555 & 941.115 & 3221.358 & 8511.047 & --        & Reject  & 0.772073 \\
		12 & 4.410 & 39.115 & 156.014 & 339.641 & 1145.220 & 5230.967 & 12710.487 & --       & Reject  & 0.772073 \\
		13 & 3.261 & 35.498 & 90.042 & 245.210 & 1239.478 & 4803.654 & 11794.881 & --       & Reject  & 0.772073 \\
		14 & 4.812 & 40.464 & 156.266 & 235.115 & 1190.418 & 4464.485 & 11196.546 & --       & Reject  & 0.772073 \\
		15 & 3.252 & 33.014 & 105.268 & 295.230 & 1191.910 & 4344.718 & 10958.238 & 0.772598 & Accept  & 0.772598 \\
		16 & 3.873 & 26.300 & 101.130 & 323.598 & 1013.381 & 5020.555 & 11903.667 & --       & Reject  & 0.772598 \\
		17 & 2.362 & 34.093 & 80.883 & 312.901 & 1111.551 & 3994.904 & 10157.245 & --       & Reject  & 0.772598 \\
		18 & 3.224 & 24.264 & 85.429 & 264.461 & 983.159 & 3048.948 & 8089.718 & --        & Reject  & 0.772598 \\
		19 & 3.881 & 34.090 & 115.527 & 213.399 & 951.002 & 3994.217 & 9745.323 & --        & Reject  & 0.772598 \\
		20 & 2.672 & 33.956 & 94.321 & 264.157 & 1249.828 & 4953.275 & 12104.281 & --       & Reject  & 0.772598 \\
		\hline
	\end{tabular}
	\caption{The first several steps of the local search for $m=6$.}
\label{tab:local-search}
\end{table}

\section{Lower Bound for Five Machines}

In this section, we give a direct construction of an instance showing that the lower bound for online scheduling on five identical machines is at least $1.74916$.

Define
\begin{equation}\label{f}
	f(x) = 16x^{4} - 46x^{3} - 32x^{2} - 13x + 42.	
\end{equation}
For $0 < x < 1$,
\[
f'(x) = 64x^{3} - 138x^{2} - 64x - 13 < -74x^{2} - 64x - 13 < 0.
\]
Therefore, $f(x)$ is monotonically decreasing on $(0, 1)$. Note that
$$f\left( \frac{7491}{10000} \right) = \frac{4154052220311}{625000000000000} > 0$$ and $$f\left( \frac{7492}{10000} \right) = -\frac{21983208093}{4882812500000} < 0,$$
It follows that $f(x)$ has a unique root $\beta$ in $(0,1)$, which satisfies
\begin{equation}\label{r6}
	\frac{7491}{10000} < \beta < \frac{7492}{10000}.
\end{equation}
More precisely, $\beta \approx 0.7491596$.

In the following, we define some quantities related to $\beta$. We then state a lemma providing several equalities and inequalities among these quantities, whose proof is deferred to the appendix.

Let
\begin{align}
	s_1 &= 15\beta(2\beta-1)\approx 5.59981,	\label{s1}\\
	s_0 &= \frac{6-4\beta}{5\beta} s_1=3(2\beta-1)(6-4\beta)=-24\beta^2 + 48\beta - 18\approx 4.48990, 	\label{s0}\\
	s_2 &=\frac{1}{2\beta-1}(s_0+s_1)= \frac{6+\beta}{5\beta(2\beta-1)} s_1=3(6+\beta)=3\beta+18\approx 20.24748, \label{s2}\\
	t_1 &=(2\beta+1)s_1-s_0=15\beta(2\beta-1)(2\beta+1)-(-24\beta^2 + 48\beta - 18)
	\nonumber\\
	&=60\beta^3 + 24\beta^2 - 63\beta + 18\approx 9.50021, \label{t1}\\
	t_2 & =4(s_2-s_1)-5s_0-t_1 \nonumber\\
	&=4\left(3\beta+18-15\beta(2\beta-1)\right)-5(-24\beta^2 + 48\beta - 18)-(60\beta^3 + 24\beta^2 - 63\beta + 18) \nonumber\\
	&=-60\beta^3-24\beta^2-105\beta+144\approx 26.64097, \label{t2}\\
	r_4 &= \beta s_0=\frac{6-4\beta}{5} s_1=3\beta(2\beta-1)(6-4\beta)=-24\beta^3 + 48\beta^2 - 18\beta\approx 3.36365, \label{r4}\\
	r_3& =\frac{1}{3}(2(s_2-s_1)-5r_4-t_1) \nonumber\\
	&=\frac{1}{3} \left(2\left(3\beta+18-15\beta(2\beta-1)\right)-5(-24\beta^3 + 48\beta^2 - 18\beta)-(60\beta^3 + 24\beta^2 - 63\beta + 18)\right) \nonumber\\
	&=20\beta^3-108\beta^2+63\beta+6\approx 0.99229, \label{r3}\\	
	r_2 &=\frac{1}{5}s_1-r_3=3\beta(2\beta-1)-(20\beta^3-108\beta^2+63\beta+6)\nonumber\\
	&=-20\beta^3+114\beta^2-66\beta-6\approx 0.12767, \label{r2}\\	
	r_1 &=s_0-r_2 - r_3-r_4=s_0-\frac{1}{5}s_1-r_4 \nonumber\\
	&=(-24\beta^2 + 48\beta - 18)- 3\beta(2\beta-1)-(-24\beta^3 + 48\beta^2 - 18\beta)\nonumber\\
	& =24\beta^3 - 78\beta^2 + 69\beta - 18\approx 0.00629. \label{r1}
\end{align}	

\begin{lemma}\label{l0}
	(i) $\frac{r_2}{r_1+r_2} \geq \beta$.
	
	(ii) $\frac{r_3}{r_1+r_2+r_3}\geq \beta$.
	
	(iii) $2r_4+3r_3<3r_4+r_3<t_1+r_3+5r_2+5r_1<2s_1$.
	
	(iv) $s_1+s_2<5r_4+t_1$.
	
	(v) $\frac{s_0+s_1+2s_2}{s_0+4r_4+t_1} \geq 1+\beta$.
	
	(vi) $t_2+r_4+2r_3<s_2+t_1+2r_3+4r_2+4r_1<2s_1+s_2+r_3+r_2+r_1<s_1+s_2+2r_4$.
	
	(vii) $\frac{s_0+s_1+s_2+t_2}{s_1+s_2+2r_4} = 1+\beta$.
\end{lemma}

We now present the construction of the instance used to prove the lower bound for $5$ machines. While this instance shares a similar structure with the general case, the processing times of all jobs can be explicitly specified.

Let $N=6$ and $k=4$. For each $1\leq i \leq 4$, let the jobs in Phase $i$ be denoted by $R_i$ with processing time $r_i$. In the next two phases, there are $4$ jobs with the same processing time and one job with a larger processing time. For $i = 5, 6$, let $S_{i-4}$ and $T_{i-4}$ denote the jobs with smaller and larger processing times in Phase $i$, respectively, with processing times $s_{i-4}$ and $t_{i-4}$. Finally, let $F$ denote the last job, with processing time $2s_2$.

Analogous to Lemma \ref{l1}, we demonstrate that for the competitive ratio of the algorithm to be less than $1+\beta$, all jobs within the same phase must be assigned to distinct machines. Specifically, for each $1\leq i\leq 4$, if algorithm $A$ assigns the jobs within each of the first $i-1$ phases to distinct machines but fails to assign the jobs in Phase $i$ to distinct machines, then $\frac{C^A(\mathcal{J}_{i1})}{C^*(\mathcal{J}_{i1})} \geq 1+\beta$. For each $5\leq i\leq 6$, if algorithm $A$ assigns the jobs within each of the first $i-1$ phases to distinct machines but fails to assign the smaller jobs in Phase $i$ to distinct machines, then $\frac{C^A(\mathcal{J}_{i1})}{C^*(\mathcal{J}_{i1})} \geq 1+\beta$. Moreover, if algorithm $A$ assigns the jobs within each of the first $i-1$ phases and the smaller jobs in Phase $i$ to distinct machines, but the larger job in Phase $i$ shares a machine with another job in the same phase, then $\frac{C^A(\mathcal{J}_{i2})}{C^*(\mathcal{J}_{i2})} \geq 1+\beta$.
Consequently, the competitive ratio of algorithm $A$ is at least $1+\beta$.

If the jobs in Phase $1$ are not assigned to distinct machines, then $C^A(\mathcal{J}_{11}) \geq 2r_1$. Obviously, $C^*(\mathcal{J}_{11}) = r_1$. Then $\frac{C^A(\mathcal{J}_{11})}{C^*(\mathcal{J}_{11})} \ge \frac{2r_1}{r_1} = 2$.

If the jobs in Phase $1$ are assigned to distinct machines, but those in Phase $2$ are not, then $C^A(\mathcal{J}_{21}) \geq r_1+2r_2$. Obviously, $C^*(\mathcal{J}_{21}) = r_1+r_2$. Then by Lemma \ref{l0}(i),
$$\frac{C^A(\mathcal{J}_{21})}{C^*(\mathcal{J}_{21})} \ge \frac{r_1+2r_2}{r_1+r_2} \geq 1+\beta.$$

If the jobs within each of the first two phases are assigned to distinct machines, but those in Phase $3$ are not, then $C^A(\mathcal{J}_{31}) \geq r_1+r_2+2r_3$. Obviously, $C^*(\mathcal{J}_{31}) = r_1+r_2+r_3$. Then by Lemma \ref{l0}(ii),
$$\frac{C^A(\mathcal{J}_{31})}{C^*(\mathcal{J}_{31})} \ge \frac{r_1+r_2+2r_3}{r_1+r_2+r_3} \geq 1+\beta.$$

If the jobs within each of the first three phases are assigned to distinct machines, but those in Phase $4$ are not, then  $C^A(\mathcal{J}_{41}) \geq r_1+r_2+r_3+2r_4$. Obviously, $C^*(\mathcal{J}_{41}) = r_1+r_2+r_3+r_4$. Then by \eqref{r1} and \eqref{r4},
$$\frac{C^A(\mathcal{J}_{41})}{C^*(\mathcal{J}_{41})} \ge \frac{r_1+r_2+r_3+2r_4}{r_1+r_2+r_3+r_4}=\frac{s_0+r_4}{s_0}= 1+\beta.$$

If the jobs within each of the first four phases are assigned to distinct machines, but those smaller jobs in Phase $5$ are not, then $C^A(\mathcal{J}_{51}) \geq s_0+2s_1$. The jobs assigned to each machine in the optimal schedule of $\mathcal{J}_{51}$ are shown as follows,
$$\{R_1, R_4, S_1\},\{R_1, R_4, S_1\}, \{R_1, R_4, S_1\}, \{R_1, R_4, S_1\}, \{R_1, R_2, R_2, R_2, R_2, R_2, R_3, R_3, R_3, R_3, R_3, R_4\}.$$
By \eqref{r2} and \eqref{r1}, $r_1+r_4+s_1=r_1+5r_2+5r_3+r_4$ and thus $C^*(\mathcal{J}_{51}) =r_1+r_4+s_1=s_0+\frac{4}{5}s_1$. Then by \eqref{s0},
$$\frac{C^A(\mathcal{J}_{51})}{C^*(\mathcal{J}_{51})} \ge \frac{s_0+2s_1}{s_0+\frac{4}{5}s_1}=\frac{\frac{6-4\beta}{5\beta}+2}{\frac{6-4\beta}{5\beta}+\frac{4}{5}}= 1+\beta.$$

If the jobs within each of the first four phases and the smaller jobs in Phase $5$ are assigned to distinct machines, but the larger job in Phase $5$ shares a machine with another job in Phase $5$, then $C^A(\mathcal{J}_{52}) \geq s_0+s_1+t_1$. The jobs assigned to each machine in a feasible schedule of $\mathcal{J}_{52}$ are shown as follows,
$$\{S_1, S_1\},\{S_1, S_1\}, \{R_3, R_4, R_4, R_4\}, \{R_3, R_3, R_3, R_4, R_4\}, \{R_1, R_1, R_1, R_1, R_1, R_2, R_2, R_2, R_2, R_2, R_3, T_1\}.$$
Thus by Lemma \ref{l0}(iii), 
$$C^*(\mathcal{J}_{52}) \leq \max\{2s_1, 3r_4+r_3, 2r_4+3r_3, t_1+r_3+5r_2+5r_1\}=2s_1.$$ 
Then by \eqref{t1},
$$\frac{C^A(\mathcal{J}_{52})}{C^*(\mathcal{J}_{52})} \ge \frac{s_0+s_1+t_1}{2s_1}=1+\beta.$$

If the jobs within each of the first five phases are assigned to distinct machines, but those smaller jobs in Phase $6$ are not, then $C^A(\mathcal{J}_{61}) \geq s_0+s_1+2s_2$. The jobs assigned to each machine in a feasible schedule of $\mathcal{J}_{61}$ are shown as follows,
$$\{R_1, R_2, R_3, S_1, S_2\},\{R_1, R_2, R_3, S_1, S_2\}, \{R_1, R_2, R_3, S_1, S_2\}, \{R_1, R_2, R_3, S_1, S_2\},$$ 
$$\{R_1, R_2, R_3, R_4, R_4, R_4, R_4, R_4, T_1\}.$$
Thus by Lemma \ref{l0}(iv) and \eqref{r1}, 
$$C^*(\mathcal{J}_{61}) \leq \max\{r_1+r_2+r_3+s_1+s_2, r_1+r_2+r_3+5r_4+t_1\}=r_1+r_2+r_3+5r_4+t_1=s_0+4r_4+t_1.$$ 
Then by Lemma \ref{l0}(v),
$$\frac{C^A(\mathcal{J}_{61})}{C^*(\mathcal{J}_{61})} \ge \frac{s_0+s_1+2s_2}{s_0+4r_4+t_1} \geq 1+\beta.$$

If the jobs within each of the first five phases and the smaller jobs in Phase $6$ are assigned to distinct machines, but the larger job in Phase $6$ shares a machine with another job in Phase $6$, then $C^A(\mathcal{J}_{62}) \geq s_0+s_1+s_2+t_2$. The jobs assigned to each machine in a feasible schedule of $\mathcal{J}_{62}$ are shown as follows,
$$\{S_1, S_2, R_4, R_4\},\{S_1, S_2, R_4, R_4\}, \{R_1, R_2, R_3, S_1, S_1, S_2\},$$
$$\{R_1, R_1, R_1, R_1, R_2, R_2, R_2, R_2, R_3, R_3, T_1, S_2\}, \{R_3, R_3, R_4, T_2\}.$$
Thus by Lemma \ref{l0}(vi),  
$$C^*(\mathcal{J}_{62})\leq \max\{s_1+s_2+2r_4, 2s_1+s_2+r_3+r_2+r_1, s_2+t_1+2r_3+4r_2+4r_1,  t_2+r_4+2r_3\}=s_1+s_2+2r_4.$$ 
Then by Lemma \ref{l0}(vii),
$$\frac{C^A(\mathcal{J}_{62})}{C^*(\mathcal{J}_{62})} \ge \frac{s_0+s_1+s_2+t_2}{s_1+s_2+2r_4} = 1+\beta.$$

Therefore, we have proved that all jobs in the same phase must be assigned to distinct machines. Finally, when the last job arrives, $C^A(\mathcal{J}) \geq s_0+s_1+s_2+2s_2$. The jobs assigned to each machine in the optimal schedule of $\mathcal{J}$ are shown as follows, 
$$\{S_2, S_2\},\{S_2, S_2\}, \{R_1, R_1, R_1, R_1, R_1, R_2, R_2, R_2, R_2, R_2, R_3, R_3, S_1, S_1, T_2\}, $$
$$\{R_3, R_3, R_3, R_4, R_4, R_4, R_4, R_4, S_1, S_1, T_1\}, \{F\}.$$
By \eqref{r3}, \eqref{r1} and \eqref{t2}, $3r_3+5r_4+2s_1+t_1=2s_2$ and
\begin{align*}
	5r_1+5r_2+2r_3+2s_1+t_2& =5(s_0-r_2 - r_3-r_4)+5r_2+2r_3+2s_1+(4(s_2-s_1)-5s_0-t_1)\\
	&=-3r_3-5r_4+4s_2-2s_1-t_1=2s_2.	
\end{align*}
Thus  $C^*(\mathcal{J}) =2s_2$. Then by \eqref{s2},
$$\frac{C^A(\mathcal{J})}{C^*(\mathcal{J})} \ge \frac{s_0+s_1+s_2+2s_2}{2s_2} =1+\beta.$$

Based on the above discussion, we obtain the following theorem.

\begin{theorem}\label{thm3}
	There is no online algorithm with competitive ratio smaller than
	$1+\beta \approx 1.7491596$ for scheduling on five parallel machines.
\end{theorem}

The method presented in the previous section can also be applied to the case of five machines. However, it does not yield a lower bound better than that in Theorem~\ref{thm3}. This is because the instance in Theorem~\ref{thm3}, despite its limited number of job types, is carefully constructed so that the optimal value can be bounded more tightly. As the number of machines increases, characterizing the optimum becomes extremely difficult, which highlights the limitation of the direct construction approach.

\section{Conclusions}

In this paper, we present new lower bounds for the online scheduling problem on $m\geq 5$ parallel machines. For $m=5$, we construct the adversarial instances directly. For larger values of $m$, we estimate the optimal makespan by exploiting algorithms for the bin packing problem, which enables us to generate the corresponding instances efficiently via computer search. Each approach has its own advantages. The direct construction yields tighter estimates of the optimal value and thus leads to stronger lower bounds. However, it relies on problem-specific insights, ad hoc techniques, and delicate numerical relationships among the parameters involved. It is therefore difficult to draw general conclusions, and the computational effort becomes prohibitively large as the number of machines increases.

The lower bound we establish not only provides a targeted guideline for designing optimal algorithms, but also suggests new directions for improving the lower bound further. Nevertheless, despite the improvements achieved, we do not believe these lower bounds are tight. To strengthen them further, the most direct approach would be to obtain tighter and more accurate estimates of the optimal makespan. Ultimately, however, new instance structures may be required. A major difficulty is that the structure of instances may depend on the number of machines. In addition, it remains open whether instances with a finite number of jobs can yield lower bounds that match the competitive ratios of online algorithms.

\section{Appendix}

\subsection{Proof of Lemma \ref{l0}}

\begin{proof}
	Write $h = \frac{3}{4} - \beta$, $a = \frac{8}{10000}$ and $b = \frac{9}{10000}$. By \eqref{r6}, $0 < a < h < b$.
	
	By \eqref{r1}, \eqref{r2},  \eqref{r4} and \eqref{s0}, we have
	\begin{equation}\label{r5}
		r_1 + r_2 = (24\beta^3 - 78\beta^2 + 69\beta - 18)+(-20\beta^3+114\beta^2-66\beta-6)=4\beta^{3}+36\beta^{2}+3\beta-24
	\end{equation}
	and
	\begin{equation}\label{r0}
		r_1 + r_2 + r_3 = s_0 - r_4 = (1 - \beta)s_0=(1 - \beta)(-24\beta^2 + 48\beta - 18)=24\beta^{3}-72\beta^{2}+66\beta-18.
	\end{equation}

	(i) By \eqref{r2} and \eqref{r5},
	\begin{align*}
		r_2 - \beta(r_1 + r_2)	&=-20\beta^3+114\beta^2-66\beta-6-\beta(4\beta^{3}+36\beta^{2}+3\beta-24)\\
		&= -4\beta^{4} - 56\beta^{3} + 111\beta^{2} - 42\beta - 6 \\
		&= -4\left( \frac{3}{4} - h \right)^{4} - 56\left( \frac{3}{4} - h \right)^{3} + 111\left( \frac{3}{4} - h \right)^{2} - 42\left( \frac{3}{4} - h \right) - 6 \\
		&= -4h^{4}+68h^{3}-\frac{57}{2}h^{2}-\frac{93}{4}h+\frac{3}{64}
		> -4b^{4}+68a^{3}-\frac{57}{2}b^{2}-\frac{93}{4}b+\frac{3}{64} \\
		& =\frac{64817374533439}{2500000000000000} > 0.
	\end{align*}
	Hence, $\frac{r_2}{r_1+r_2} \geq \beta$. 
	
	(ii) By \eqref{r3} and \eqref{r0},
	\begin{align*}
		r_3 - \beta(r_1 + r_2 + r_3) 
		&=20\beta^3-108\beta^2+63\beta+6-\beta(24\beta^{3}-72\beta^{2}+66\beta-18)\\
		&= -24\beta^{4} + 92\beta^{3} - 174\beta^{2} + 81\beta + 6 \\
		&= -24\left( \frac{3}{4} - h \right)^{4} + 92\left( \frac{3}{4} - h \right)^{3} - 174\left( \frac{3}{4} - h \right)^{2} + 81\left( \frac{3}{4} - h \right) + 6 \\
		&= -24h^{4}-20h^{3}-48h^{2}+\frac{261}{4}h+\frac{3}{32}
		> -24b^{4}-20b^{3}-48b^{2}+\frac{261}{4}a+\frac{3}{32}\\ 
		&= \frac{182388881755317}{1250000000000000} > 0.
	\end{align*}
	Hence, $\frac{r_3}{r_1+r_2+r_3}\geq \beta$.
	
	(iii) By \eqref{r4} and  \eqref{r3},
	\begin{align*}
		r_4 - 2r_3
		&= (-24\beta^3 + 48\beta^2 - 18\beta) - 2\left( 20\beta^{3} - 108\beta^{2} + 63\beta + 6 \right) = -64\beta^{3} + 264\beta^{2} - 144\beta - 12 \\
		&= -64\left( \frac{3}{4} - h \right)^{3} + 264\left( \frac{3}{4} - h \right)^{2} - 144\left( \frac{3}{4} - h \right) - 12 \\
		&= 64h^{3}+120h^{2}-144h+\frac{3}{2}> 64a^{3}+120a^{2}-144b+\frac{3}{2} =\frac{669178141}{488281250} > 0.
	\end{align*}
	By \eqref{t1}, \eqref{r5} and \eqref{r4},
	\begin{align*}
		t_1+5r_2+5r_1 -3r_4&=60\beta^3 + 24\beta^2 - 63\beta + 18+5(4\beta^{3}+36\beta^{2}+3\beta-24)-3(-24\beta^3 + 48\beta^2 - 18\beta)\\
		&= 152\beta^{3}+60\beta^{2}+6\beta-102 \\
		&= 152\left( \frac{3}{4} - h \right)^{3}+60\left( \frac{3}{4} - h \right)^{2}+6\left( \frac{3}{4} - h \right)-102\\
		&=-152h^{3}+402h^{2}-\frac{705}{2}h+\frac{3}{8} > -152b^{3}+402a^{2}-\frac{705}{2}b+\frac{3}{8} \\
		&=\frac{7250896149}{125000000000} > 0.
	\end{align*}
	By \eqref{s1}, \eqref{t1}, \eqref{r3} and \eqref{r5},
	\begin{align*}
		2s_1 - t_1 - r_3 - 5r_2 - 5r_1&=30\beta(2\beta-1)-(60\beta^3 + 24\beta^2 - 63\beta + 18)\\
		& \quad -(20\beta^3-108\beta^2+63\beta+6)-5(4\beta^{3}+36\beta^{2}+3\beta-24)\\
		&= -100\beta^{3} - 36\beta^{2} - 45\beta + 96 \\
		&= -100\left( \frac{3}{4} - h \right)^{3} - 36\left( \frac{3}{4} - h \right)^{2} - 45\left( \frac{3}{4} - h \right) + 96 \\
		&= 100h^{3}-261h^{2}+\frac{1071}{4}h-\frac{3}{16} > 100a^{3}-261b^{2}+\frac{1071}{4}a-\frac{3}{16}\\
		&=\frac{66221603}{2500000000} > 0.
	\end{align*}
	Hence, $2r_4+3r_3<3r_4+r_3<t_1+r_3+5r_2+5r_1<2s_1$.
	
	(iv) By \eqref{r4}, \eqref{t1}, \eqref{s1} and \eqref{s2},
	\begin{align*}
		5r_4 + t_1 - s_1 - s_2
		&= 5(-24\beta^3 + 48\beta^2 - 18\beta)+(60\beta^3 + 24\beta^2 - 63\beta + 18)-15\beta(2\beta-1)-(3\beta+18)\\
		&= -60\beta^{3} + 234\beta^{2} - 141\beta = -60\left( \frac{3}{4} - h \right)^{3} + 234\left( \frac{3}{4} - h \right)^{2} - 141\left( \frac{3}{4} - h \right) \\
		&= 60h^{3}+99h^{2}-\frac{435}{4}h+\frac{9}{16} > 60a^{3}+99a^{2}-\frac{435}{4}b+\frac{9}{16}  =\frac{1452151221}{3125000000} > 0.
	\end{align*}
	Hence, $s_1+s_2<5r_4+t_1$. 
	
	(v) By \eqref{s0}, \eqref{s1},  \eqref{s2}, \eqref{r4}  and \eqref{t1},
	\begin{align*}
		s_0 + s_1 + 2s_2 - (1 + \beta)(s_0 + 4r_4 + t_1)& =-\beta s_0 + s_1 + 2s_2 - (1 + \beta)(4r_4+t_1) \\
		&=-\beta(-24\beta^2 + 48\beta - 18)+15\beta(2\beta-1)+2(3\beta+18)\\
		& \quad - (1 + \beta)(4(-24\beta^3 + 48\beta^2 - 18\beta)+(60\beta^3 + 24\beta^2 - 63\beta + 18)) 
		\\
		&= 36\beta^{4} - 156\beta^{3} - 99\beta^{2} + 126\beta + 18 \\
		&= 36\left( \frac{3}{4} - h \right)^{4} - 156\left( \frac{3}{4} - h \right)^{3} - 99\left( \frac{3}{4} - h \right)^{2} + 126\left( \frac{3}{4} - h \right) + 18 \\
		&= 36h^{4}+48h^{3}-\frac{657}{2}h^{2}+225h+\frac{153}{64} \\
		&> 36a^{4}+48a^{3}-\frac{657}{2}b^{2}+225a+\frac{153}{64} =\frac{200809292155527}{78125000000000} > 0.
	\end{align*}
	Hence, $\frac{s_0+s_1+2s_2}{s_0+4r_4+t_1} \geq 1+\beta$.
	
	(vi) By \eqref{s2}, \eqref{t1},  \eqref{t2}, \eqref{r5} and \eqref{r4},  
	\begin{align*}
		s_2+t_1-t_2+4r_2+4r_1-r_4 & =(3\beta+18)+(60\beta^3 + 24\beta^2 - 63\beta + 18)-(-60\beta^3-24\beta^2-105\beta+144) \\
		& \quad +4(4\beta^{3}+36\beta^{2}+3\beta-24)-(-24\beta^3 + 48\beta^2 - 18\beta) \\
		& =160\beta^{3}+144\beta^{2}+75\beta-204 \\
		& =160\left( \frac{3}{4} - h \right)^{3}+144\left( \frac{3}{4} - h \right)^{2}+75\left( \frac{3}{4} - h \right)-204 \\
		& =-160h^{3}+504h^{2}-561h+\frac{3}{4}>-160b^{3}+504a^{2}-561b+\frac{3}{4}\\
		&=\frac{1533890271}{6250000000}>0.
	\end{align*}
	By \eqref{s1}, \eqref{t1},   \eqref{r3} and  \eqref{r5},	
	\begin{align*}
		2s_1-t_1-r_3-3r_1-3r_2 &= 30\beta(2\beta-1)-(60\beta^3 + 24\beta^2 - 63\beta + 18) \\
		& \quad -(20\beta^{3} - 108\beta^{2} + 63\beta + 6)-3(4\beta^{3}+36\beta^{2}+3\beta-24) \\
		&=-92\beta^{3}+36\beta^{2}-39\beta+48 \\
		&=-92\left( \frac{3}{4} - h \right)^{3}+36\left( \frac{3}{4} - h \right)^{2}-39\left( \frac{3}{4} - h \right)+48 \\
		&=92h^3 - 171h^2 + \frac{561}{4}h + \frac{3}{16}>92a^3 - 171b^2 + \frac{561}{4}a + \frac{3}{16} \\
		&=\frac{18722596069}{62500000000}>0.
	\end{align*}
	By \eqref{r4}, \eqref{s1}   and  \eqref{r0},
	\begin{align*}
		2r_4 - s_1 - r_3 - r_2 - r_1&= 2(-24\beta^3 + 48\beta^2 - 18\beta) - 15\beta(2\beta-1) -(24\beta^{3}-72\beta^{2}+66\beta-18)  \\
		&= -72\beta^{3} + 138\beta^{2} - 87\beta + 18 \\
		&= -72\left( \frac{3}{4} - h \right)^{3} + 138\left( \frac{3}{4} - h \right)^{2} - 87\left( \frac{3}{4} - h \right) + 18 \\
		&= 72h^{3}-24h^{2}+\frac{3}{2}h > 72a^{3}-24b^{2}+\frac{3}{2}a =\frac{9223413}{7812500000} > 0.
	\end{align*}
	Hence, $t_2+r_4+2r_3<s_2+t_1+2r_3+4r_2+4r_1<2s_1+s_2+r_3+r_2+r_1<s_1+s_2+2r_4$. 
	
	(vii) By \eqref{s0}, \eqref{s1},  \eqref{s2}, \eqref{t2} and  \eqref{r4},
	\begin{align*}
		s_0 + s_1 + s_2 + t_2 - (1 + \beta)(s_1 + s_2 + 2r_4)&=s_0-\beta s_1 - \beta s_2+t_2-2(1 + \beta)r_4 \\
		&= (-24\beta^2 + 48\beta - 18)-\beta \cdot 15\beta(2\beta-1) - 3\beta(\beta + 6) \\
		&\quad + \left( -60\beta^3-24\beta^2-105\beta+144 \right) \\
		&\quad -2(1 + \beta)(-24\beta^3 + 48\beta^2 - 18\beta) \\
		&= 48\beta^{4} - 138\beta^{3} - 96\beta^{2} - 39\beta + 126 = 3f(\beta) = 0.
	\end{align*}
\end{proof}

\end{document}